# On the Mixed $H_2/H_\infty$ Loop Shaping Trade-offs in Fractional Order Control of the AVR System

Saptarshi Das and Indranil Pan

***Abstract*—This paper looks at frequency domain design of a fractional order (FO) PID controller for an Automatic Voltage Regulator (AVR) system. Various performance criteria of the AVR system are formulated as system norms and is then coupled with an evolutionary multi-objective optimization (MOO) algorithm to yield Pareto optimal design trade-offs. The conflicting performance measures consist of the mixed $H_2/H_\infty$ designs for objectives like set-point tracking, load disturbance and noise rejection, controller effort and as such are an exhaustive study of various conflicting design objectives. A fuzzy logic based mechanism is used to identify the best compromise solution on the Pareto fronts. The advantages and disadvantages of using a FOPID controller over the conventional PID controller, which are popular for industrial use, are enunciated from the presented simulations. The relevance and impact of FO controller design from the perspective of the dynamics of AVR control loop is also discussed.**

## I. Introduction

APPLICATION of fractional calculus based control system designs have gained impetus in recent times due to the flexibility and effectiveness that can be gained through such methodologies [1-2]. These controllers use the FO integro-differential operators as extra tuning knobs which can be tuned to meet additional design constraints and provide additional robustness to the design [3-4]. Merging the concepts of computational intelligence techniques with FO control designs are also being recently explored with encouraging results [5].

However applications of fractional order controllers for electrical power and energy systems are still largely unexplored. A few studies have been done for the application of the FOPID controller to the design of the AVR in a power system. In [6-7], the FOPID parameters are tuned for an AVR system with swarm based optimization algorithms which show better time domain performance over that with conventional PID structure. In [8] a multi-objective formalism has been developed and the time domain performance trade-offs of various design objectives have been investigated. The obtained results are mixed, with the PID performing better under some objectives and the FOPID performing better at others. However, all these investigations suffer from a common problem. The objective functions are framed in time domain and the evolutionary algorithms minimize some time domain performance index. Therefore, the obtained results give no indication of other important design focuses like robust stability, disturbance rejection capability etc. To alleviate these issues, the problems in power system control can be framed in frequency domain and the system performance can be expressed in terms of system norms which need to be minimized or maximized. In [9], the FOPID design for an AVR system has been done in frequency domain. The conflicting objectives being maximization of phase margin (for better oscillation damping) and gain crossover frequency (for higher speed of operation) are studied by coupling them with multi-objective evolutionary algorithms. In this paper, we extend this concept to an exhaustive set of frequency domain design criteria for the AVR system and show the achievable set of performances with the FOPID controller structure.

The rest of the paper is organized as follows. Section II briefly introduces the theoretical background of the problem with the basics of FO control using system norms. Section III shows the design trade-offs for several combinations of conflicting objectives and compares the best compromise solutions for each controller structure. The paper ends with the conclusion as section IV.

## II. Theoretical Background of Fractional Order Control Design for AVR System

### *A. Fractional Calculus Based Control Systems*

Fractional calculus extends the common notion of integer order integration or differentiation to any arbitrary real number. It can be represented by ${}_aD_t^\alpha$ where $\alpha \in \Re$ is the order of the differentiation or integration. There are many definitions of fractional calculus like the Grünwald-Letnikov (GL), Riemann-Liouville (RL) and Caputo definitions [1-2]. In this paper, the Caputo definition is used for realizing the fractional integro-differential operators of the FOPID controller. According to Caputo's definition, the $\alpha^{th}$ order derivative of a function *f(t)* with respect to time is given by (1)

S. Das is with Communications, Signal Processing and Control (CSPC) Group, School of Electronics and Computer Science, University of Southampton, Southampton SO17 1BJ, UK (e-mail: s.das@soton.ac.uk).

I. Pan is with Energy, Environment, Modelling and Minerals ($E^2M^2$) Research Section, Department of Earth Science and Engineering, Imperial College London, Exhibition Road, SW7 2AZ, UK (email: i.pan11@imperial.ac.uk).

$$D^{\alpha} f(t) = \frac{1}{\Gamma(m-\alpha)} \int_0^t \frac{D^m f(t)}{(t-\tau)^{\alpha+1-m}} d\tau, \tag{1}$$
$$\alpha \in \Re^+, m \in Z^+, m-1 \le \alpha < m$$

### B. AVR System and Controller Structure

In the conventional AVR loop in Fig. 1, there is a sensor element $H(s)$ in the feedback path. Let the power system, generator and exciter system as a whole be represented as the transfer function $G(s)$ similar to the treatments in [9] which is then to be controlled by a FOPID controller $C(s)$. If the closed loop system is represented as an equivalent system with unity feedback with the same controller, then let the effective open loop system be represented by $G_{eff}(s)$. Fig. 1 shows the equivalence of the original AVR control loop and its equivalent unity feedback representation with the same FOPID controller in (2).

$$C(s) = K_P + \left(K_i / s^{\lambda}\right) + \left(K_d s^{\mu} / \left(1 + T_f s^{\mu}\right)\right) \tag{2}$$

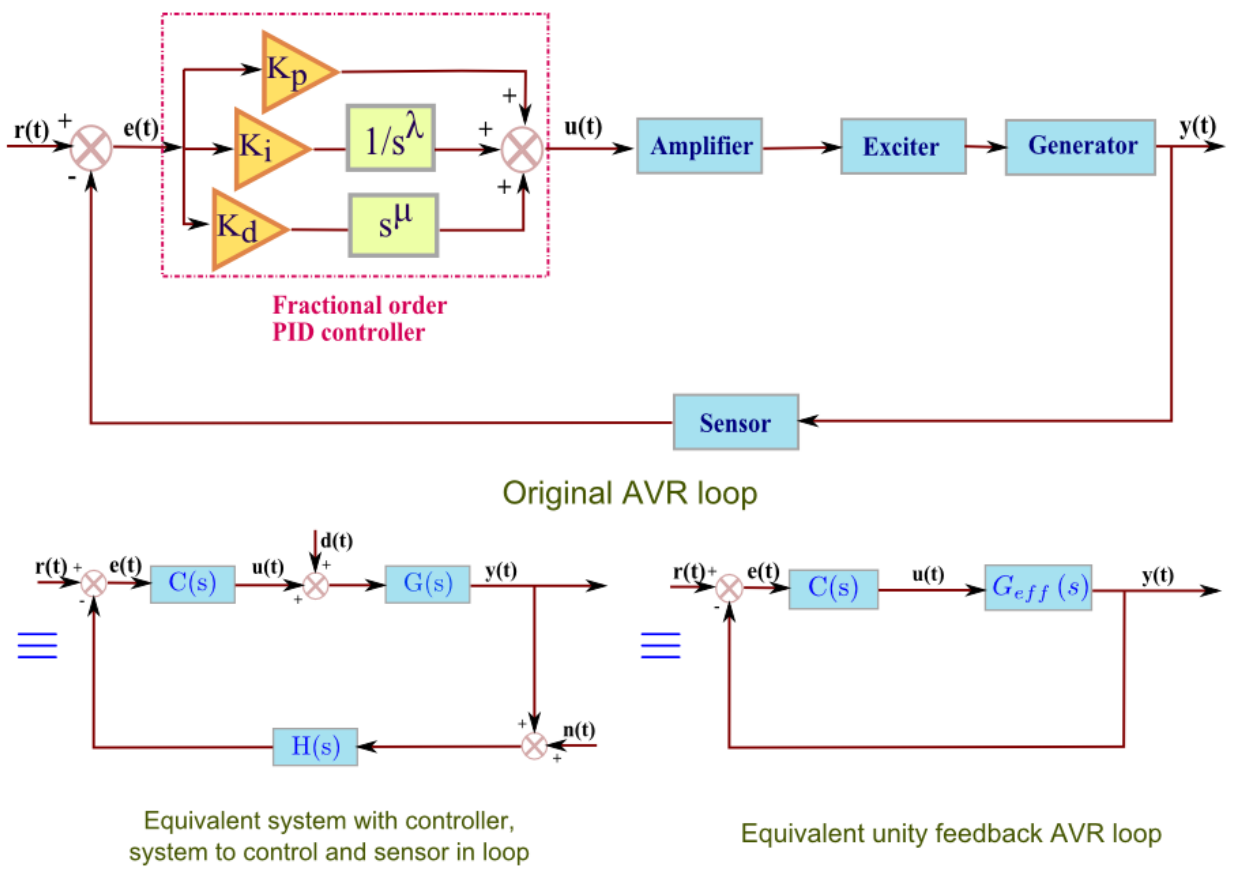

Fig. 1 AVR control loop and its equivalent unity feedback structure.

Here, the design parameters are $\{K_p, K_i, K_d, T_f, \lambda, \mu\}$ representing the proportional, integral and derivative gains, derivative filter time constant and integro-differential orders respectively. Here, a filtered version of the derivative action is considered over the conventional or unfiltered structure of FOPID controller in [8-10] since in calculating most of the system/signal norms in loop shaping approach, the nine system transfer functions (which maps the set-point, load disturbance and sensor noise to the tracking error, system input with disturbance and noise corrupted measurement for feedback) need to be proper transfer functions with more poles than zeros [11]. This implies that the plant $G(s)$, controller $C(s)$ and sensor $H(s)$ all should be proper which motivates the choice of the filtered derivative action (2), as detailed in [11]. Kakhki and Haeri [12] used the FOPID controller with filtered derivative in (2) but with equal integro-differential orders i.e. $\lambda = \mu$, instead of the popular unfiltered derivative version of the FOPID structure by Podlubny [10]. In this work, the idea is extended with a provision of choosing all the parameters independently within a multi-objective optimization framework. The performance of a PID controller with filtered derivative action is also compared, by putting $\{\lambda, \mu\} = 1$. All the fractional derivative terms in the FOPID controller structure are continuously rationalized by an equivalent higher order transfer function using Oustaloup's recursive approximation (ORA) [1].

$$s^{\alpha} \approx K \prod_{k=-N}^{N} \frac{s+\omega'_k}{s+\omega_k}, K = \omega_h^{\alpha}, \omega_k = \omega_b \left(\omega_h / \omega_b\right)^{\frac{(k+N+(1+\alpha)/2)}{(2N+1)}}, \tag{3}$$
$$\omega'_k = \omega_b \left(\omega_h / \omega_b\right)^{(k+N+(1-\alpha)/2)/(2N+1)}$$

Thus, the error signal $e(t)$ can be passed through the differ-integrator (3) and the output of the system can be regarded as an approximation to the fractionally differentiated or integrated error signal $D^{\alpha} e(t)$ which is then weighted with appropriate gains to yield the control signal $u(t)$. In (3), $\alpha$ is the order of the differ-integration, $(2N+1)$ is the order of the filter and $(\omega_b, \omega_h)$ is the expected fitting range. In the present study, 5$^{th}$ order ORA has been adopted to represent the integro-differential operators of the controller within a frequency band of $\omega \in \{10^{-4}, 10^4\}$ rad/sec. Various analog and digital realization techniques have been suggested to implement FOPID controllers in real hardware. Each FO operator in the FOPID controller could be implemented in the form of higher order Finite Impulse Response (FIR) or Infinite Impulse Response (IIR) filters which show constant phase characteristics within wide range of frequency spectra [1-2]. Depending on the realization scheme and the order of approximation, the response of these filters might be different and the performance of the controller might change substantially which needs to be considered during hardware implementation of such controllers.

From Fig. 1, the two equivalent systems can be equated to find out the expression for the effective open loop system $G_{eff}(s)$ with unity feedback and the same controller $C(s)$.

$$\frac{C(s)G(s)}{1+C(s)G(s)H(s)} = \frac{C(s)G_{eff}(s)}{1+C(s)G_{eff}(s)} \tag{4}$$
$$\Rightarrow G_{eff}(s) = G(s) / \left(1 + C(s)G(s)H(s) - C(s)G(s)\right)$$

It is clear that the effective system to be controlled $G_{eff}$ contains the controller whose parameters are the design variables. Therefore, within the controller design procedure a stability checking criterion has been kept using Matlab's *isstable()* function involving higher order approximation of the closed loop transfer function as also done in [13-14], otherwise inappropriate choice of the controller parameters might make the whole system unstable.

### C. Conflicting Objectives in Loop Shaping Approach

Let us consider that the open-loop system $L(s)$ with the effective system and controller is represented by (5).

$$L(s) = C(s) G_{eff}(s) \tag{5}$$

Here, the sensitivity $S(s)$ and complementary sensitivity $T(s)$

functions are given by (6).

$$S(s)=\frac{1}{1+L(s)}=G_{er}(s), T(s)=\frac{L(s)}{1+L(s)}=G_{yr}(s)=G_{yn}(s) \quad (6)$$

The disturbance sensitivity $S_d(s)$ and control sensitivity $S_u(s)$ functions are given by (7) and (8) respectively.

$$S_d(s)=\left(G(s)/1+L(s)\right)=G_{yd}(s) \quad (7)$$

$$S_u(s)=\left(C(s)/\left(1+L(s)\right)\right)=G_{ur}(s)=G_{un}(s) \quad (8)$$

The above four transfer functions play a big role in obtaining an optimum performance of the AVR control loop. In order to frame the controller tuning task in a trade-off design template, from the above transfer functions few objective functions (9)-(12) are derived as suggested in [15, 16]. System norm based multi-objective design trade-offs for PID controller has been previously explored by Herreros *et al.* [16]. Sensitivity and complementary sensitivity peaks are used as design objectives for FOPI controllers in Chen *et al.* [17]. Other mixed $H_2/H_\infty$ loop shaping approaches of PID controller tuning can be found in [18-20]. The fundamental improvement of the present work over the above discussed literatures is that it is the first attempt to look at the merits of FOPID controllers vis-à-vis PID controllers, with different range of integro-differential orders for comparing the trade-offs between several mixed $H_2/H_\infty$ design objectives.

The disturbance rejection performance ($J_d$) is given by the $H_\infty$-norm of disturbance sensitivity function subjected to a unit step disturbance input as shown in (9).

$$J_d=\left\|(1/s)G_{yd}(s)\right\|_\infty \quad (9)$$

The control activity function ($J_u$) is the second important objective which can be represented as the $H_\infty$-norm of the control sensitivity function in (10).

$$J_u=\left\|G_{ur}(s)\right\|_\infty=\left\|G_{un}(s)\right\|_\infty=\left\|S_u(s)\right\|_\infty=\left\|C(s)/\left(1+L(s)\right)\right\|_\infty \quad (10)$$

The combined infinity norm of the weighted sensitivity and complementary sensitivity function is given by (11).

$$J_{ST}=\left\|W_S(s)S(s)\right\|_\infty+\left\|W_T(s)T(s)\right\|_\infty \quad (11)$$

Here, $W_S(s)$ and $W_T(s)$ are low pass filters to shape the sensitivity and co-sensitivity functions which controls robust stability and noise rejection performances respectively [17].

The set-point tracking performance ($J_{track}$) is given by the $H_2$-norm of the sensitivity function subjected to a unit step set-point change as shown in (12).

$$J_{track}=\left\|(1/s)G_{er}(s)\right\|_2=\left\|(1/s)\cdot\left(1/\left(1+L(s)\right)\right)\right\|_2 \quad (12)$$

The above mentioned $H_2/H_\infty$ norms (9)-(12) for different sensitivity transfer functions in (6)-(8) are defined in (13).

$$\begin{aligned}\left\|H(s)\right\|_2&=\sqrt{(1/2\pi)\int_{-\infty}^{\infty}Trace\left[H(j\omega)^T H(j\omega)\right]d\omega},\\ \left\|H(s)\right\|_\infty&=\max_\omega\left|H(j\omega)\right|\end{aligned} \quad (13)$$

The $H_\infty$ norms in (9)-(11) denote the peak gain of the corresponding frequency responses. Also the $H_2$ norm in (12) becomes finite when the system is stable and has a low pass characteristic i.e. Bode magnitude curve is drooping in nature as $\omega\to\infty$. Pioneering works on $H_2/H_\infty$ norm minimization based FO controller design has been reported in [21]-[24].

From the above mentioned four objectives, six set of design trade-offs are formed by taking two conflicting objectives simultaneously for minimization *viz.* $J_d-J_{ST}$, $J_d-J_{track}$, $J_d-J_u$, $J_{ST}-J_{track}$, $J_u-J_{ST}$, $J_u-J_{track}$. Also, another four set of design trade-offs can be formed by taking three conflicting objectives at a time viz. $J_d-J_u-J_{ST}$, $J_d-J_u-J_{track}$, $J_u-J_{ST}-J_{track}$, $J_d-J_{ST}-J_{track}$. Since, the Oustaloup's rational approximation technique is not valid for FO higher than one, the optimization is divided in four combinations of the integro-differential orders viz. $\{\lambda,\mu\}<1$ ($FOPID_1$), $\lambda>1,\mu<1$ ($FOPID_2$), $\lambda<1,\mu>1$ ($FOPID_3$), $\{\lambda,\mu\}>1$ ($FOPID_4$). Amongst the four combinations, only the last two combinations with integral order higher than one yielded stable solutions with the open loop system being a proper transfer function [11]. For the controller structure (2) with filtered derivative (compared to unfiltered version in [8-10]) and $\lambda<1$ ($FOPID_1$ and $FOPID_3$), the rationalized higher order controller and consequently the open loop system become improper transfer functions and thus some of the system norms become infinite. Therefore, the stabilizing controllers resulting in finite system norms are only reported in the simulations (i.e. $FOPID_2$ and $FOPID_4$). The weighting matrices for the sensitivity/co-sensitivity shaping problem have been chosen in (14) as suggested by Hung *et al.* [25].

$$W_S(s)=\frac{0.25s+0.025}{s^2+0.4s+1000}, W_T(s)=\frac{0.0125s^2+1.2025s+1.25}{s^2+20s+100} \quad (14)$$

### D. Description of the Performance Metrics

This section presents a brief description of the effects of various frequency domain metrics on the control system design. This is important in discerning the improvements that the FOPID controller offers over its integer order counterpart. The $T(s)$ in (6), is the closed loop transfer function from $r(t)$ to $y(t)$. The relative perturbation in $T(s)$ due to a relative perturbation in $G_{eff}(s)$ is the transfer function $S(s)$, i.e.

$$\lim_{\Delta G_{eff}\to 0}\left[\left(\Delta T/T\right)/\left(\Delta G_{eff}/G_{eff}\right)\right]=S \quad (15)$$

Therefore $S(s)$ is a measure of the sensitivity of the closed loop transfer function, to variations in the effective plant transfer function, and hence the terminology. In this sense, $S(s)$ is an indicator of robust stability of the AVR system.

Also, it can be seen that $S(s)$ is the transfer function from $r(t)$ to $e(t)$. Thus an ideal design would be one where $|S(j\omega)|$ is small over the range of frequencies of $r(t)$. Another important fact is that the peak magnitude of $S(s)$ is the reciprocal of the stability margin [11]. Assuming that $\omega_1$ is the maximum frequency of $r(t)$, $\varepsilon(\varepsilon<1)$ is the maximum allowable relative tracking error and $M_S(M_S>1)$ is the maximum value of $|S(j\omega)|$. Then for the input $r(t)=\cos(\omega t),\omega\le\omega_1$, the steady state error $|e(t)|\le\varepsilon$ and the stability margin is $1/M_S$.

In the case of the present AVR loop-shaping problem, the $G_{eff}(s)$ in (4) contains the transfer function of the controller $C(s)$ as well. The rationalization of the FO controller (2) will result in a minimum phase representation of $G_{eff}(s)$. If the controller is an integer order PID, then $G_{eff}(s)$ would have a zero at the center of the complex $s$-plane. For a minimum phase transfer function, it is possible to achieve an arbitrarily good tracking performance over a finite frequency range, while maintaining a given stability margin $1/M_S$. In other words, if $M_S>1$ and $\omega_1>0$ are pre-specified, it is always possible to make $\varepsilon$ arbitrarily small. However if $G_{eff}(s)$ is non-minimum phase, there is a limit to the achievable performance which is known as the water-bed effect [11]. In such situations, if the loop shaping controller is so designed that it pushes down $|S(j\omega)|$ in a particular frequency band, then it pops up at some other frequency range like a water-bed.

In transforming $|G_{eff}(s)|$ to $|G_{eff}(s)C(s)|$ using the FOPID controller, care should be taken so that the roll-off of $|G_{eff}(s)C(s)|$ is not very sharp near the gain cross-over. Incase this happens, $\arg(G_{eff}(s)C(s))$ would be too large near the crossover frequency and might result in negative phase margin and hence instability. This is automatically taken care of in the multi-objective optimization framework, by using a penalty function approach for the unstable solutions. Also, there should not be any pole-zero cancellation, as otherwise the system may have BIBO stability but loose internal stability. The internal stability is guaranteed if the characteristic polynomial has no root in the right-hand $s$-plane as shown in (16). Here, the system, controller and sensor transfer functions have been represented as ratio of polynomials representing the numerator ($N$) and denominator ($D$) i.e. $G=N_G/D_G$, $C=N_C/D_C$, $H=N_H/D_H$ [11].

$$\mathrm{Re}\left(\mathrm{Roots}\left(N_G N_C N_H + D_G D_C D_H\right)\right)\le 0 \tag{16}$$

The maximum sensitivity constrained designs and $H_\infty$ loop-shaping approaches of FO controllers have been previously studied in [17], [26] and [27] respectively.

### *E. Description of the Multi-objective Optimization Algorithm*

The Non-dominated Sorting Genetic Algorithm–II (NSGA-II) is used in this paper to optimize the conflicting frequency domain design objectives [9] while searching for the six parameters ($N_x$) of the controller (2) within a range of $\{K_p,K_i,K_d,T_f\}\in[0,10]$ representing the gains and time constant of derivative filter and $\{\lambda,\mu\}\in[0,2]$ for the integro-differential orders. The algorithm mimics the evolutionary process and uses crossover, mutation and reproduction operators to refine the solution in each generation. In the MOO, the population size ($>15N_x$), maximum number of generation ($>200N_x$) and the Pareto fraction has been chosen as 100, 1200 and 0.7 respectively [8]-[9]. The crossover fraction is set as 0.8 and the mutation fraction is set as 0.2 as in similar problems [8, 9].

The NSGA-II produces a set of non-dominated solutions which constitute the Pareto front. These solutions are Pareto optimal or non-dominated, in the sense that it is not possible to find another solution which gives better performance in all the objectives simultaneously. Therefore, if any other solution gives better performance in one objective, then it would definitely give a worse performance in the other objectives. The Pareto optimal solutions represent the design trade-offs in the problem itself and the designer has to choose the best compromise solution depending on his design constraints.

## III. Simulation and Results

### *A. Trade-off Designs with Different Combinations of Conflicting Objectives*

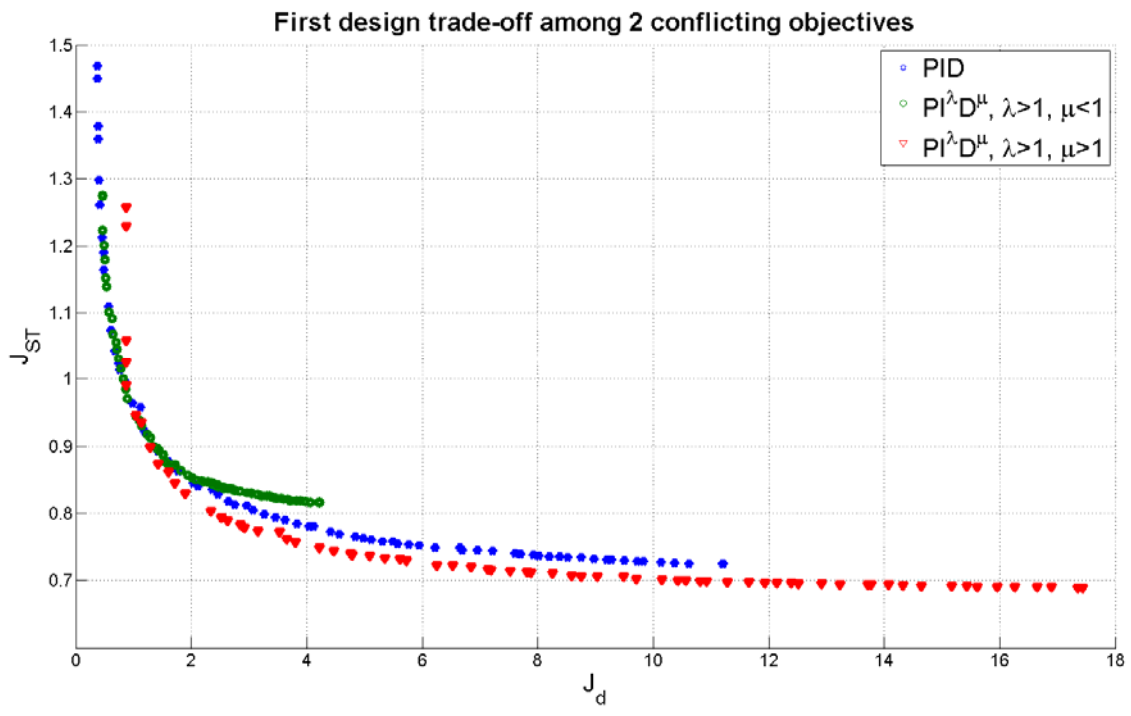

Fig. 2 First design trade-off among two objectives: $J_d$-$J_{ST}$ (Case 1).

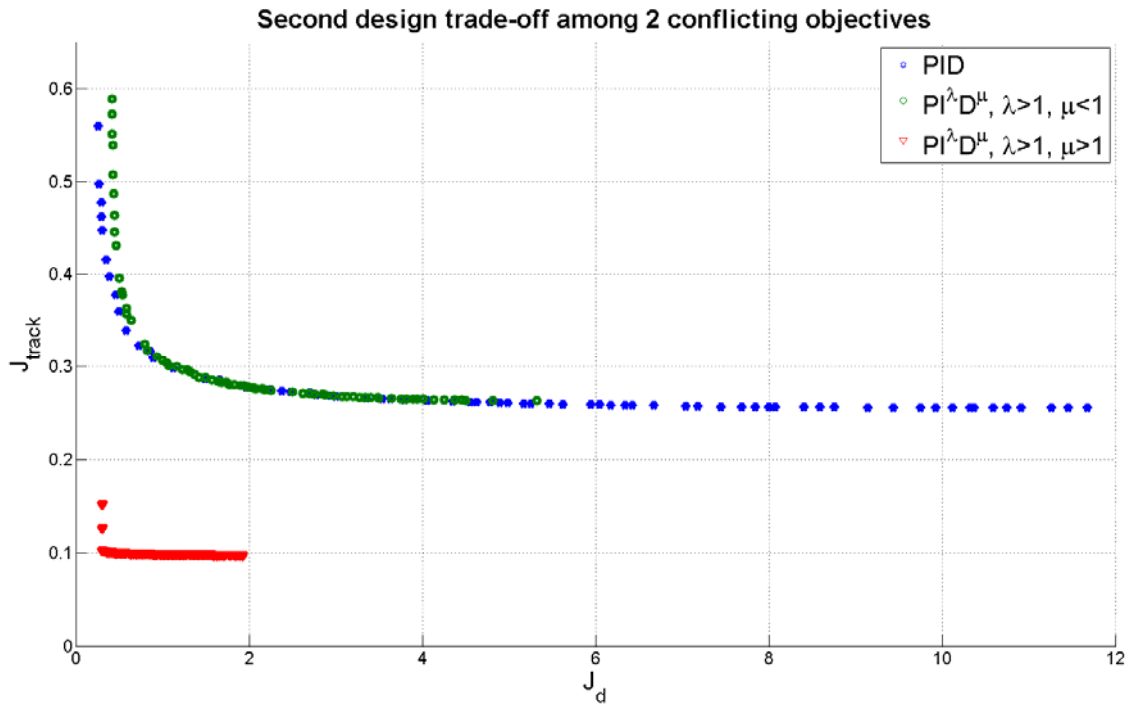

Fig. 3 Second design trade-off among two objectives: $J_d$-$J_{track}$ (Case 2).

The Pareto optimal fronts for two conflicting objectives are reported first where in all the six design trade-offs, as shown in Fig. 2-Fig. 7, the FOPID$_4$ controller with $\lambda>1,\mu>1$ gives either a better spread of the Pareto solutions or gives better non-dominated solutions than the other two Pareto fronts. This is also the case for the Pareto fronts of the sets of three conflicting objectives taken at a time, as shown in Fig. 8-Fig. 11. The figures also confirm the potential conflict that is present in control design for simultaneously obtaining low values of all the objectives (9)-(12).

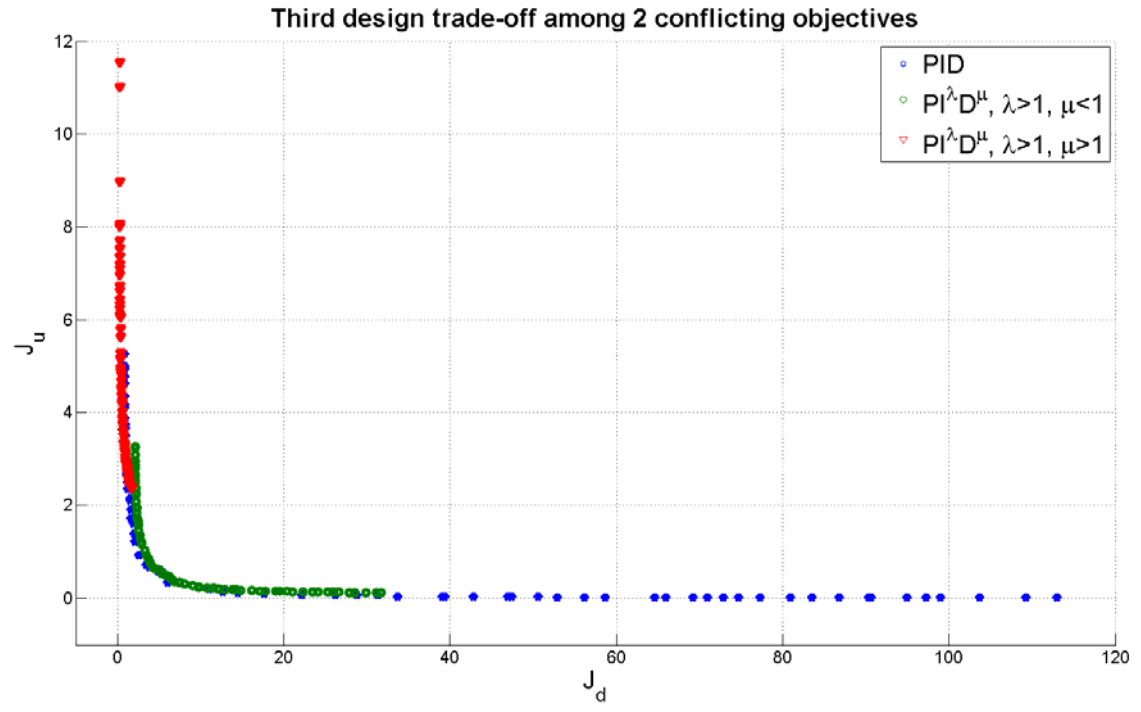

Fig. 4 Third design trade-off among two objectives: $J_d$-$J_u$ (Case 3).

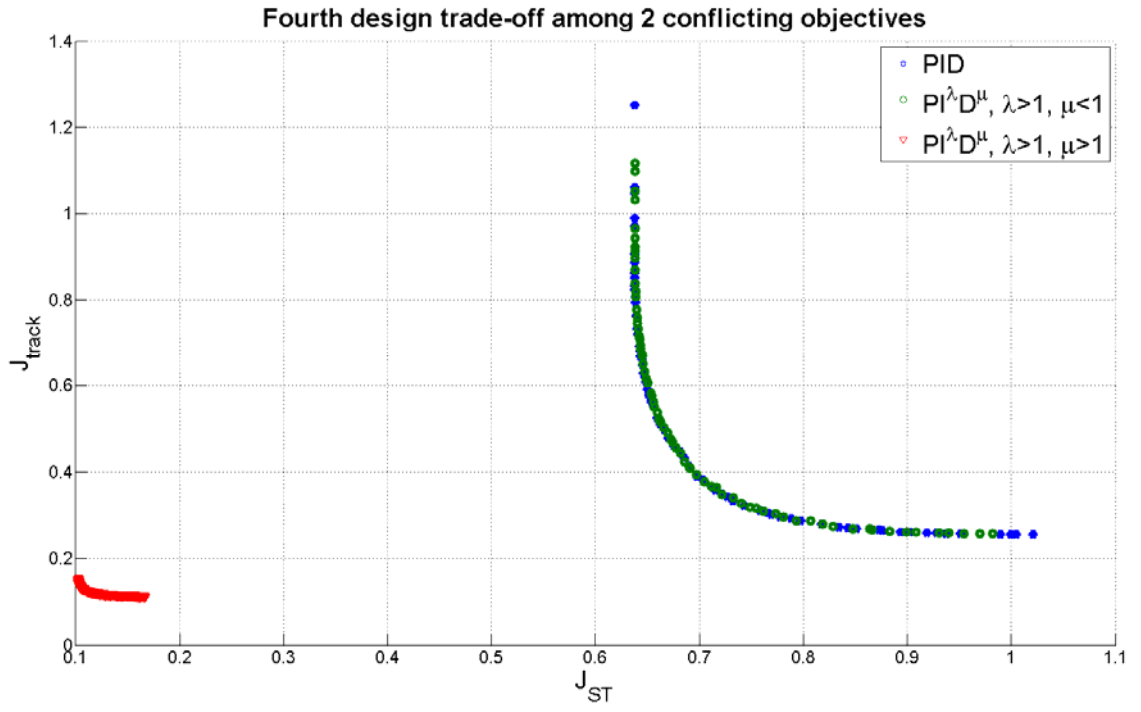

Fig. 5 Fourth design trade-off among two objectives: $J_{ST}$-$J_{track}$ (Case 4).

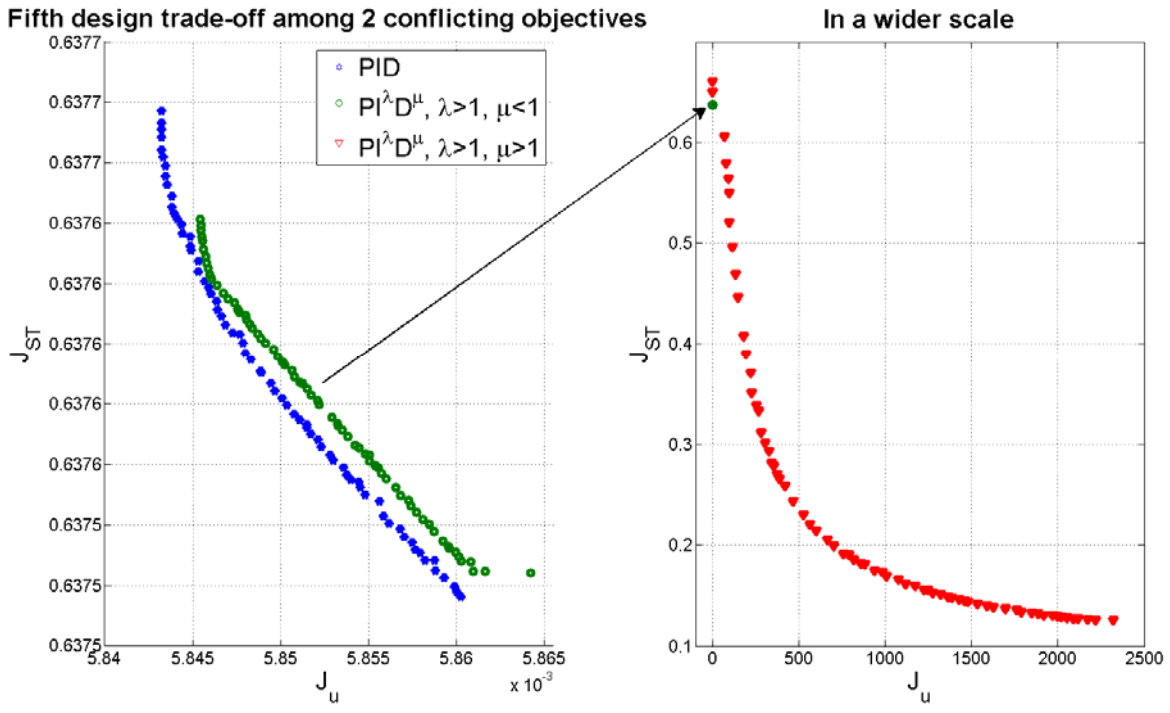

Fig. 6 Fifth design trade-off among two objectives: $J_u$-$J_{ST}$ (Case 5).

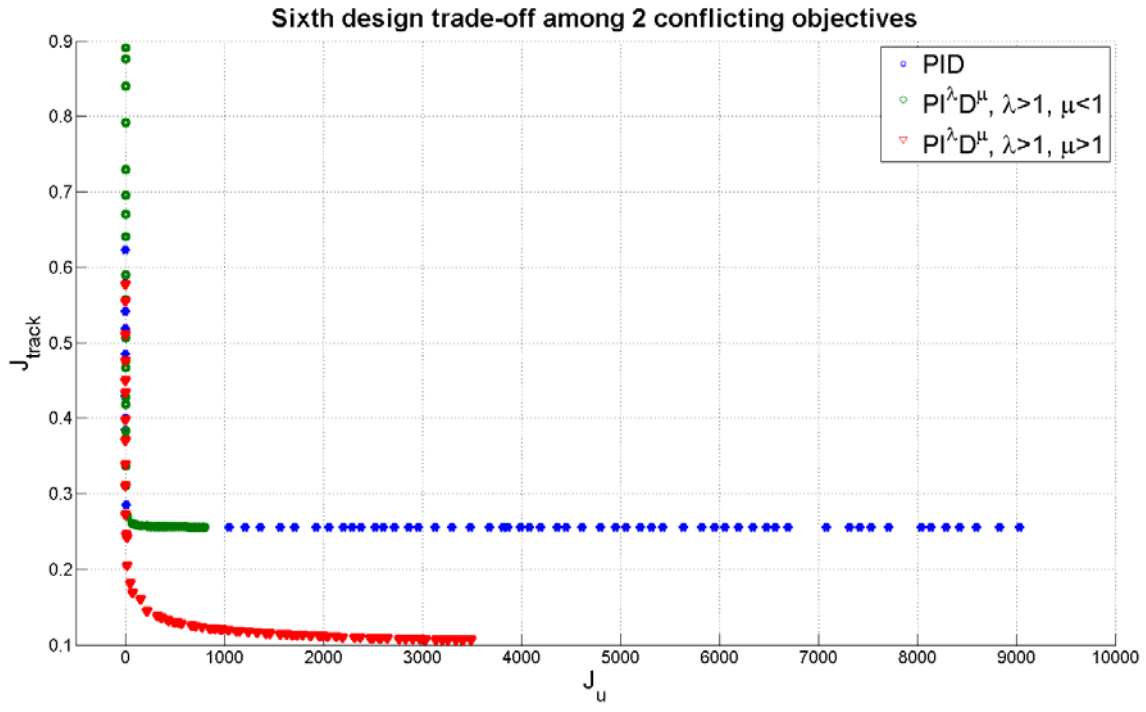

Fig. 7 Sixth design trade-off among two objectives: $J_u$-$J_{track}$ (Case 6).

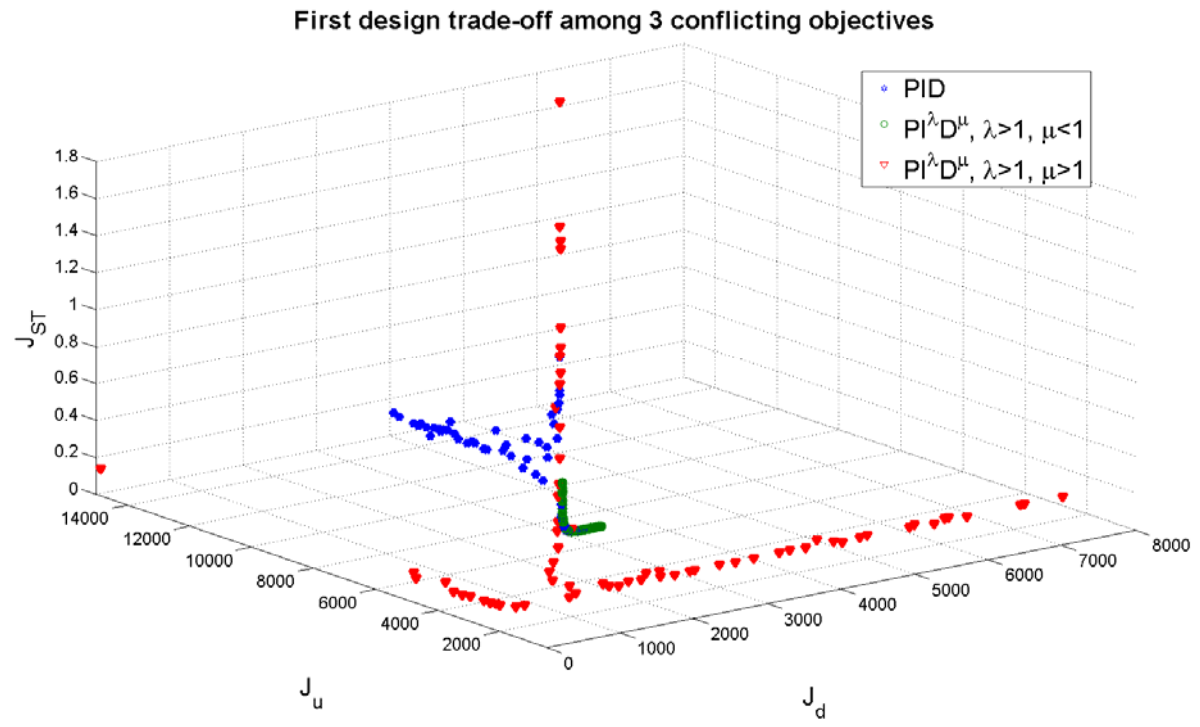

Fig. 8 First design trade-off among three objectives: $J_d$-$J_u$-$J_{ST}$ (Case 7).

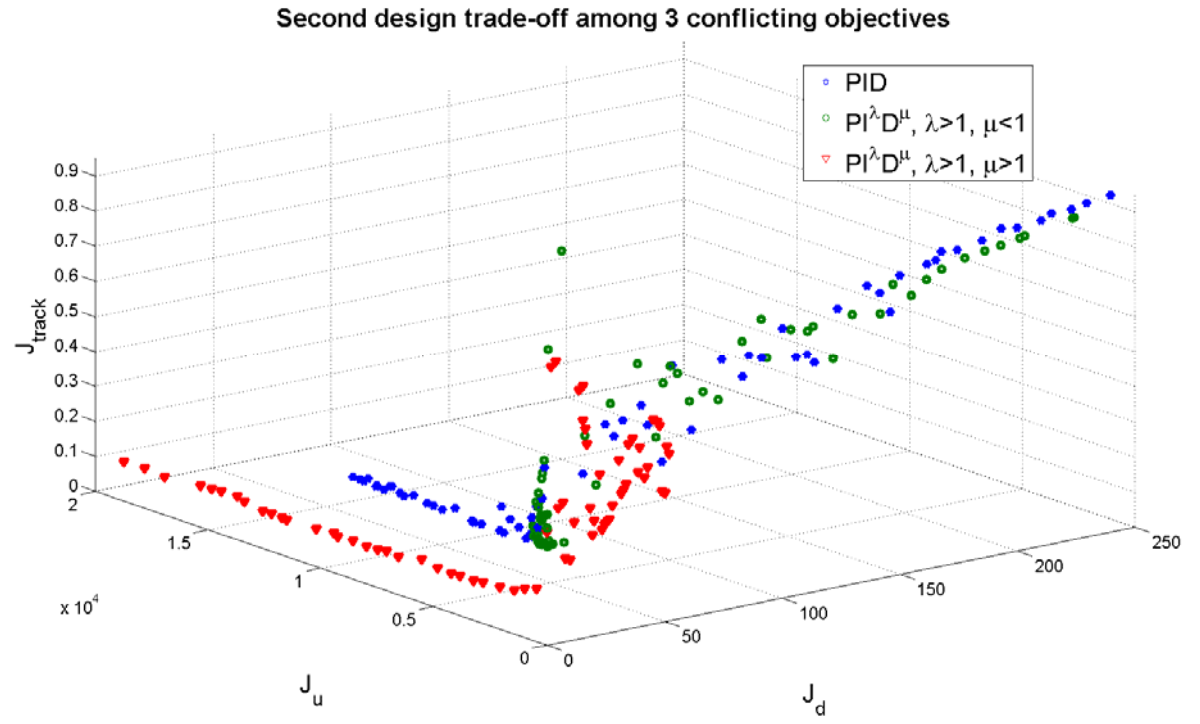

Fig. 9 Second design trade-off among three objectives: $J_d$-$J_u$-$J_{track}$ (Case 8).

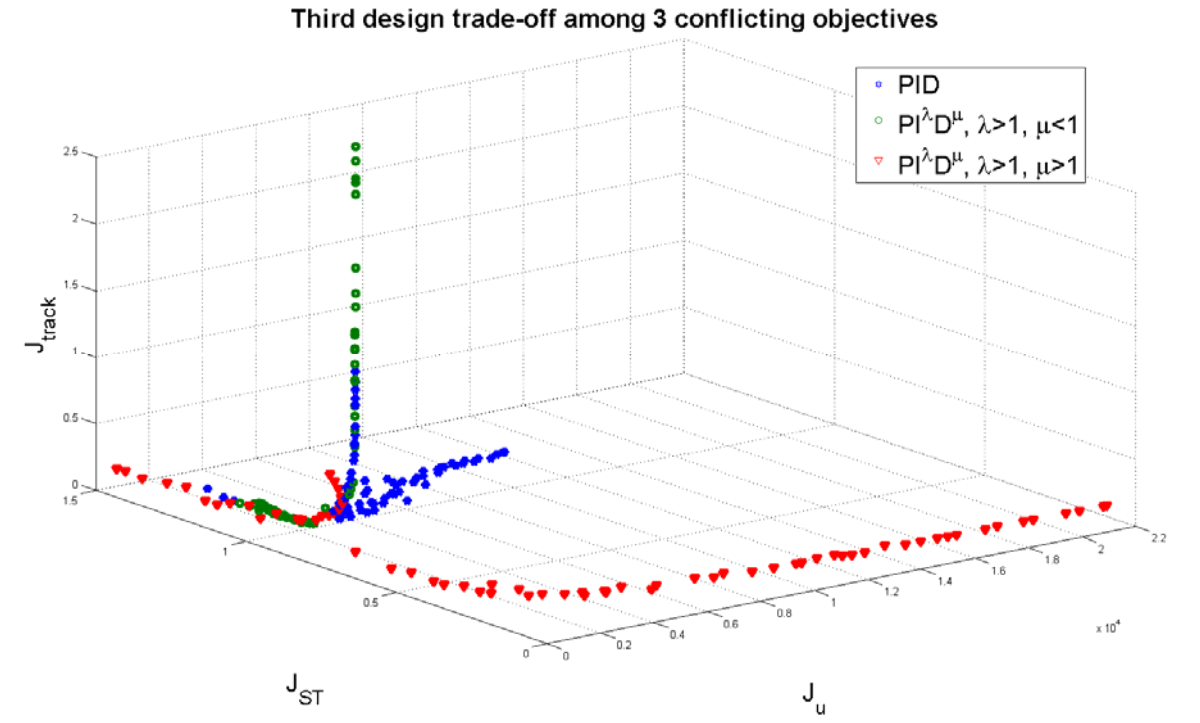

Fig. 10 Third design trade-off among three objectives: $J_u$-$J_{ST}$-$J_{track}$ (Case 9).

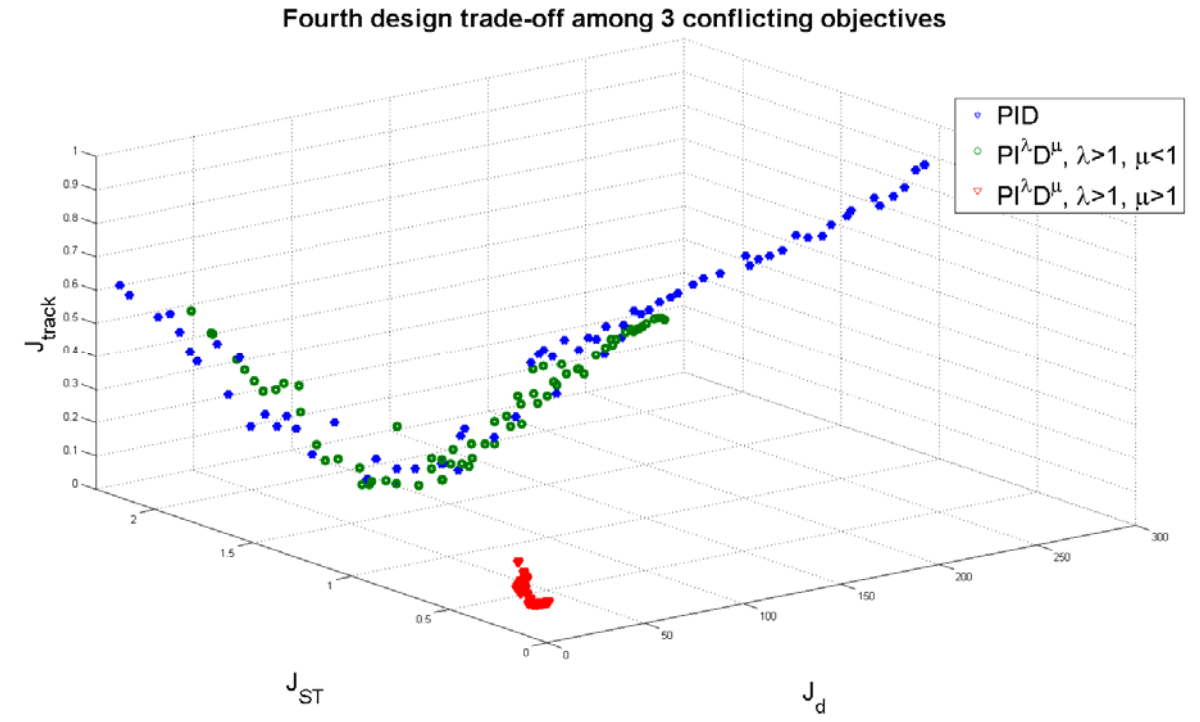

Fig. 11 Fourth design trade-off among three objectives: $J_d$-$J_{ST}$-$J_{track}$ (Case 10).

### *B. Fuzzy Logic Based Mechanism for Identifying the Best-Compromise Solutions on the Pareto Front*

The Pareto solutions as reported in the previous section are diverse in nature and it is difficult to find out a best compromise solution just by visual inspection of the Pareto trade-off curves. So a fuzzy logic based method [28] is adopted to assist in systematically choosing the best compromise solution. The designer might have imprecise goals for each objective function and this can be encoded in the form of a fuzzy membership function $\mu_F$ . In this case, $\mu_{F_i}$ for each objective function $i$ is taken to be a strictly monotonic and decreasing continuous function expressed as (17).

$$\mu_{F_i} = \begin{cases} 1 & if \quad F_i \le F_i^{\min} \\ \left(F_i^{\max} - F\right) / \left(F_i^{\max} - F_i^{\min}\right) & if \ F_i^{\min} \le F_i \le F_i^{\max} \\ 0 & if \quad F_i \ge F_i^{\max} \end{cases} \quad (17)$$

The numerical value of $\mu_{F_i}$ basically represents the degree to which a particular solution has satisfied the objective $F_i$ . It is a value between zero (worst satisfaction) to one (best satisfaction). The degree of satisfaction of the objectives by a solution can be represented by (18).

$$\mu^k = \left(\sum_{i=1}^{M} \mu_i^k\right) \Bigg/ \left(\sum_{j=1}^{N_{PS}} \sum_{i=1}^{M} \mu_i^j\right) \quad (18)$$

where, $M$ is the number of objectives and $N_{PS}$ is the number of solutions in the Pareto front. The best compromise solution in the Pareto front is the one for which (18) is maximum.

### *C. Responses of the Best Compromise Solutions*

The best compromise solutions for each case are found out using the fuzzy logic based approach for the present MOO framework (given in Table I of supplementary material) and the corresponding four plots for the four performance objectives ( $J_{ST}, J_d, J_u, J_{track}$ ) are shown and termed as subplots (a, b, c, d) respectively for each figure. The FOPID controller with $\lambda > 1, \mu < 1$ is denoted as FOPID$_2$ and the one with $\lambda > 1, \mu > 1$ as FOPID$_4$ henceforth.

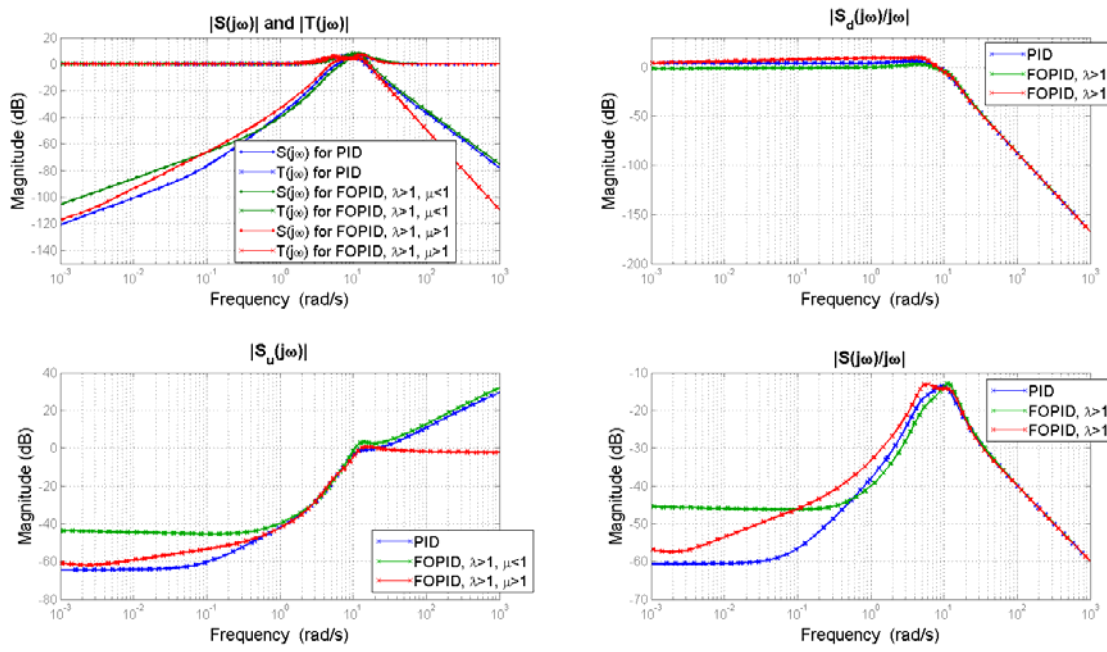

Fig. 12 Best compromise responses for $J_d$-$J_{ST}$ trade-off (Case 1).

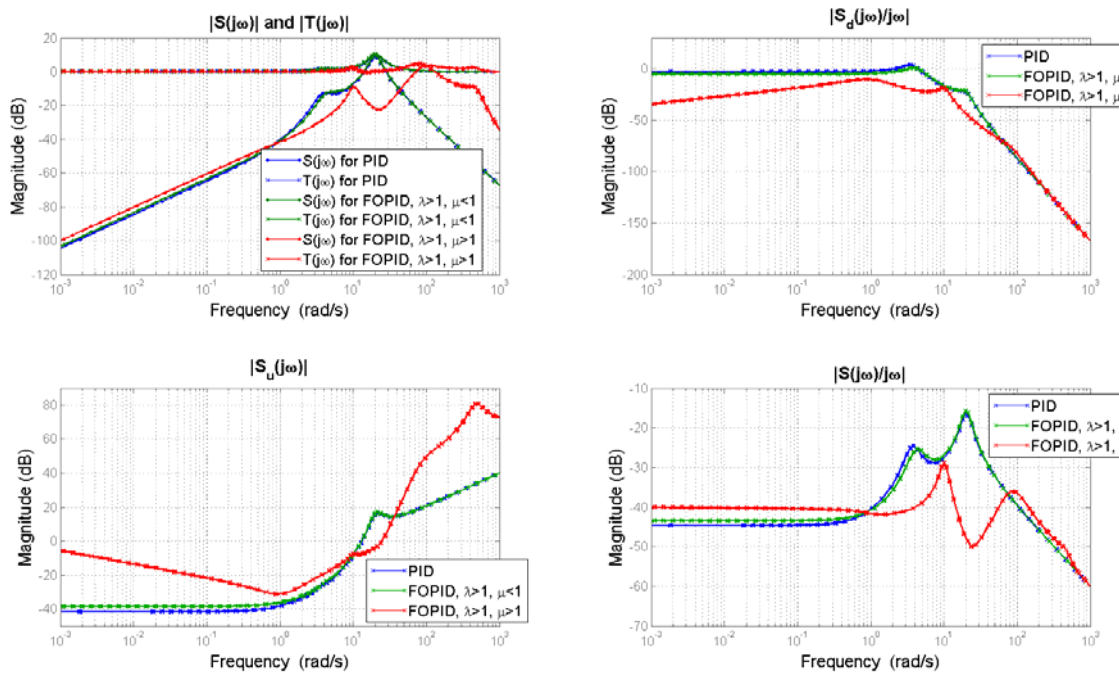

Fig. 13 Best compromise responses for $J_d$-$J_{track}$ trade-off (Case 2).

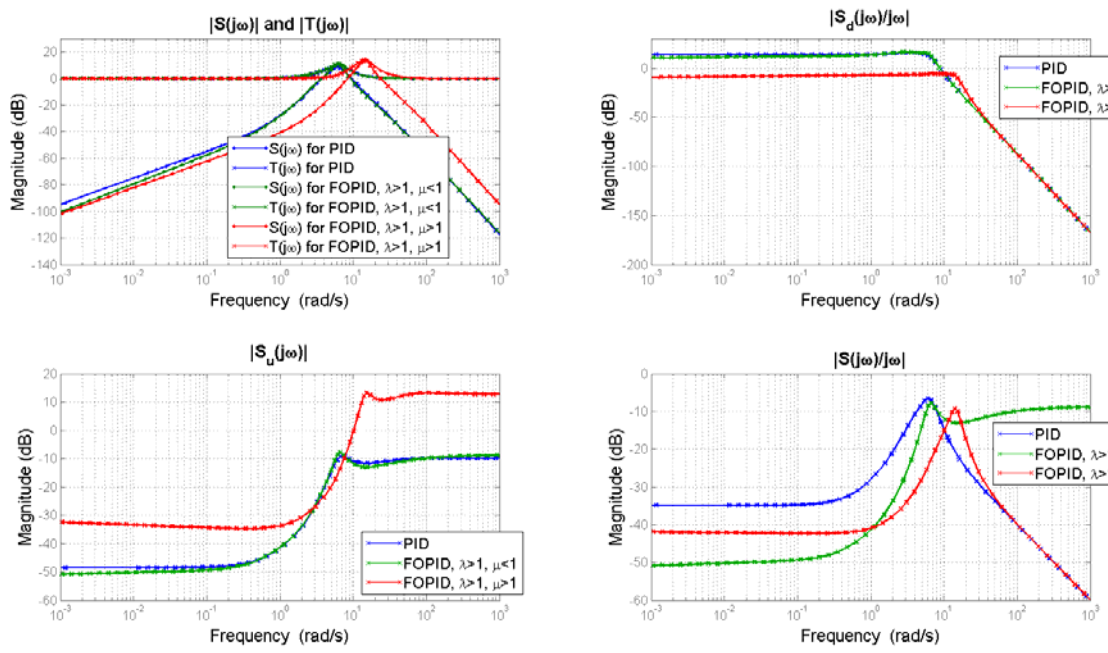

Fig. 14 Best compromise responses for $J_d$-$J_u$ trade-off (Case 3).

In Fig. 12, $J_d$ is the maximum value (infinity norm) of the magnitude plot of subplot (b) and $J_{ST}$ is the weighted sum of the maximum values of $|S(j\omega)|$ and $|T(j\omega)|$ in subplot (a). It can be observed that $J_d$ is lowest for FOPID$_2$ and highest for FOPID$_4$ and the PID case lies in between. However for $J_{ST}$ , it is the opposite. This can be verified from the Pareto fronts as well. This implies that the FOPID$_2$ would give better disturbance rejection performance but it would have a lower robust stability margin and noise rejection capability than the other two controllers. The other two subplots (c) and (d) are automatically shaped and not taken into the design criteria.

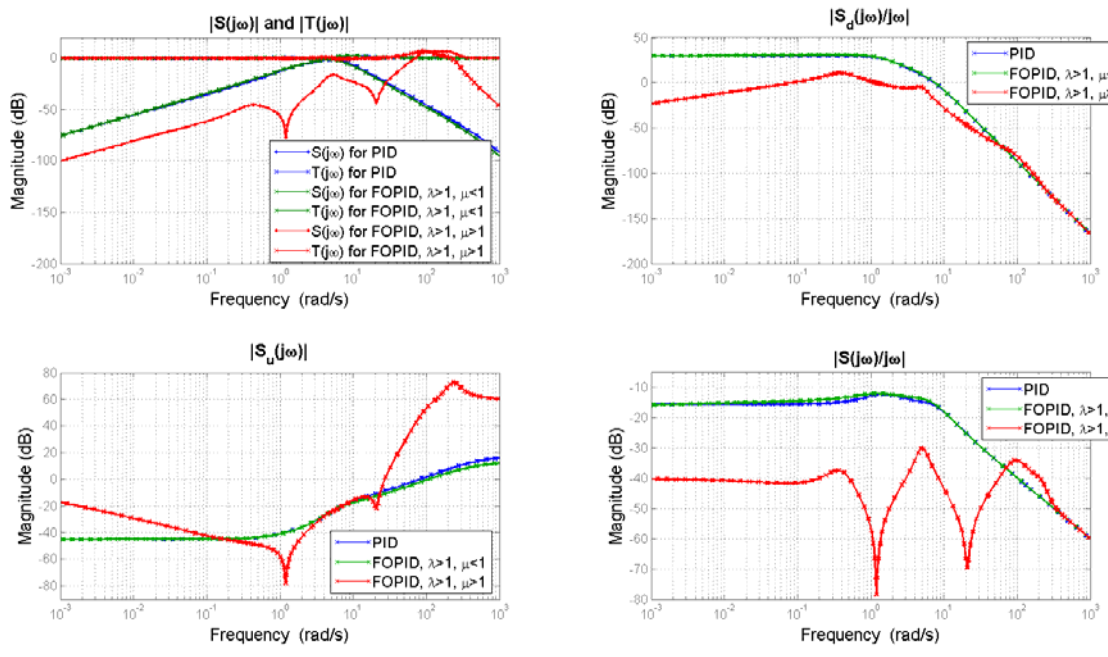

Fig. 15 Best compromise responses for $J_{ST}$-$J_{track}$ trade-off (Case 4).

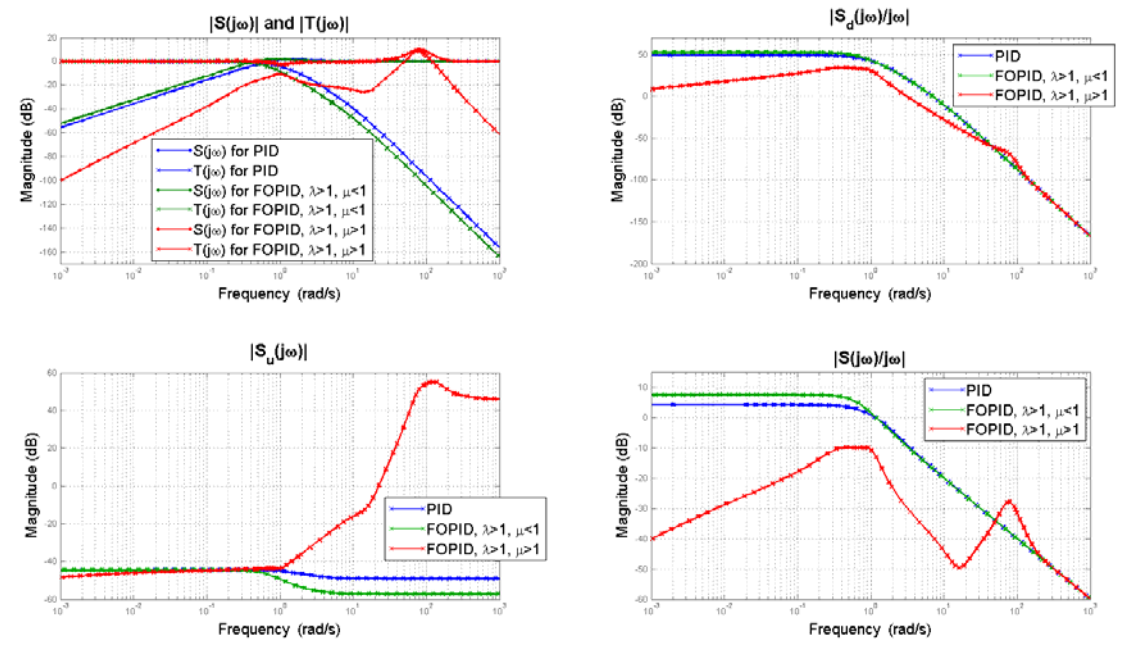

Fig. 16 Best compromise responses for $J_u$-$J_{ST}$ trade-off (Case 5).

In Fig. 13 from subplots (b) and (d) it can be seen that the FOPID$_4$ has a better disturbance rejection performance. However, the area under the curve of subplot (d) (which is the measure of the $H_2$ norm) is also the best for the FOPID$_4$ controller. Therefore the FOPID$_4$ controller structure should be chosen for such an objective as it outperforms both the other structures. A comparison can be made with Fig. 3 which shows that the Pareto for the FOPID$_4$ completely dominates the other two and hence the FOPID$_4$ is able to give better performance, simultaneously in both objectives.

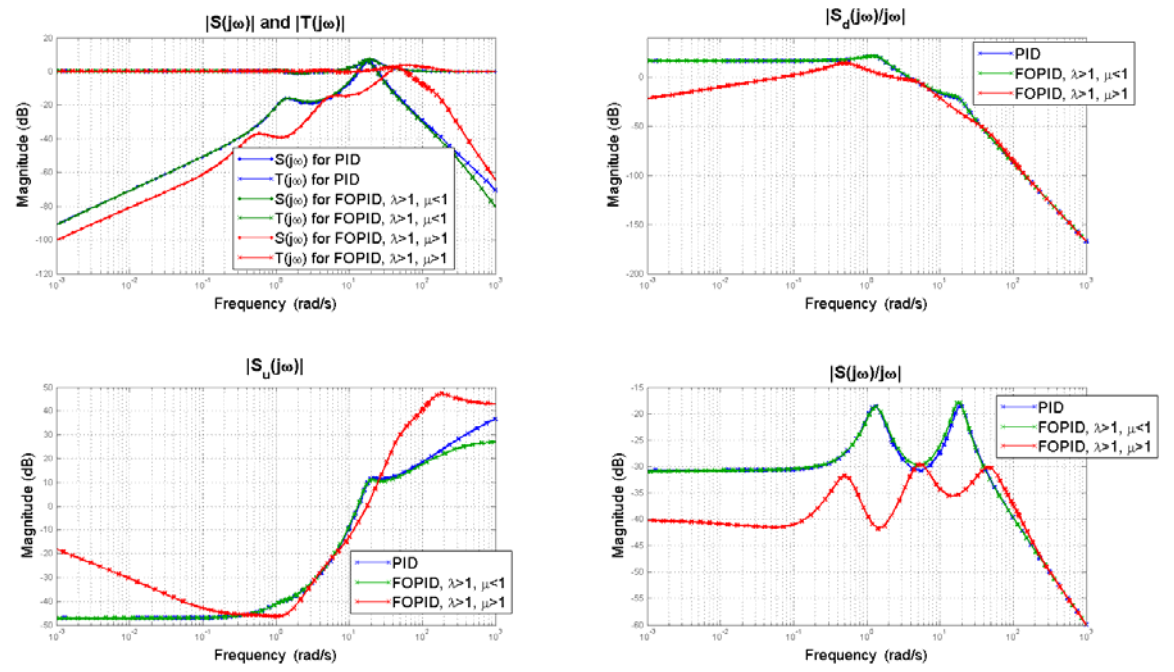

Fig. 17 Best compromise responses for $J_u$-$J_{track}$ trade-off (Case 6).

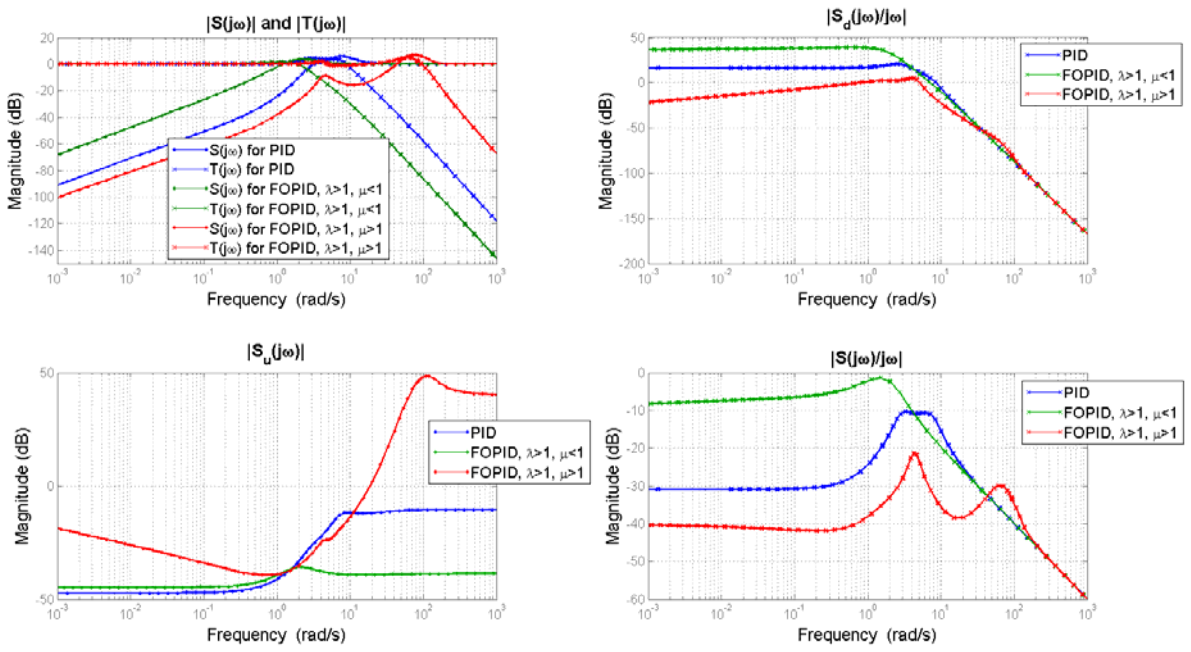

Fig. 18 Best compromise responses for $J_d$-$J_u$-$J_{ST}$ trade-off (Case 7).

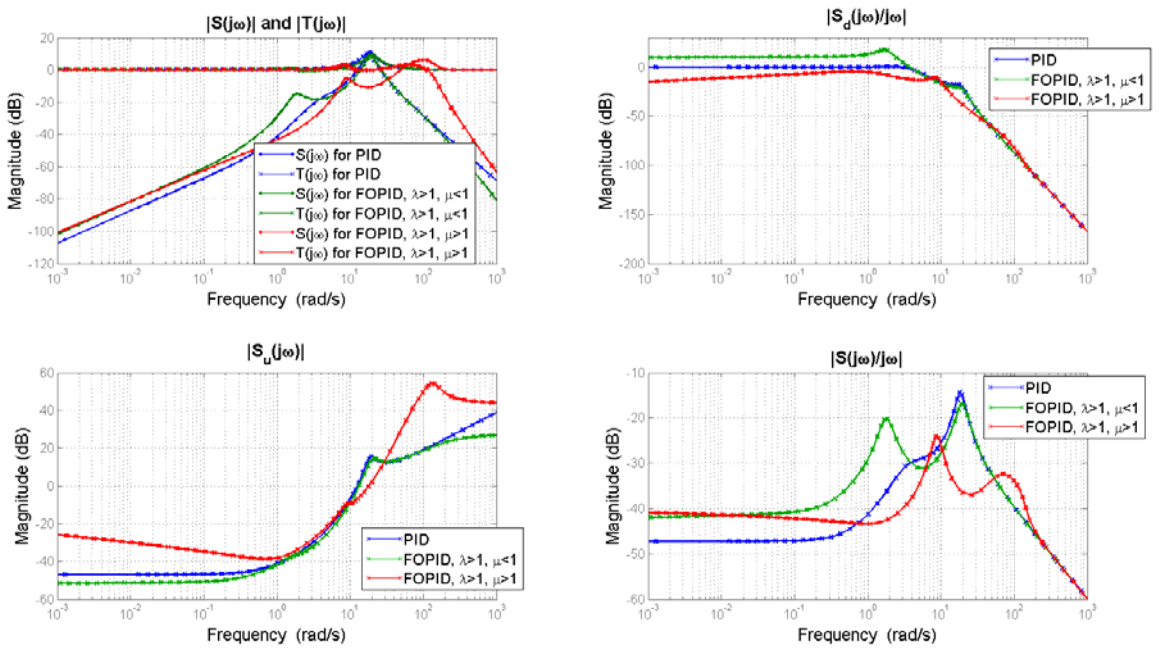

Fig. 19 Best compromise responses for $J_d$-$J_u$-$J_{track}$ trade-off (Case 8).

To understand the trade-offs in Fig. 14, sub-plots (b) and (c) must be looked into. From Fig. 14(b), the disturbance rejection capability of FOPID$_4$ is found to be higher but it takes the maximum amount of control effort as seen from Fig. 14(c). So in situations where the control effort is more expensive, the other two controller structures can be chosen at the cost of decreased disturbance rejection capability. In Fig. 15, the FOPID$_4$ structure gives a better design performance for both the objectives $J_{ST}$ and $J_{track}$ as shown in subplots (a) and (d). This is because the Pareto front of the FOPID$_4$ structure in Fig. 5, completely dominates the other two. In Fig. 16(a) the robust stability and noise rejection capabilities of the FOPID$_4$ structure is much better than the other two. However, the control signal required for FOPID$_4$ is much more as can be seen from Fig. 16(c). A closer look at the Pareto fronts for the corresponding case in Fig. 6 indicates that the FOPID$_4$ structure is capable designing controllers having a much wider range of objective function values. Therefore it is possible to obtain a controller whose control effort is much smaller, but this would imply that the corresponding robust stability and noise rejection performance would be significantly lower. In Fig. 17(c) and (d), the control effort from FOPID$_4$ is higher than the other two, but its tracking capability is better (due to smaller area under the curve of subplot (d)).

For the three objective design trade-off in Case 7, Fig. 18 (b), (c) and (a) should be compared. The FOPID$_4$ structure has the best disturbance rejection performance (as shown in Fig. 18(b)) but requires a very high amount of control effort (as shown by Fig. 18 (c)) among the three controller structures. The $J_{ST}$ criterion in Fig. 18(a) is better for FOPID$_4$ as can also be observed from the Pareto front. It should be noted that for the FOPID$_4$ structure, the sensitivity function $|S(j\omega)|$, is smaller over a wider low frequency range, implying better tracking capability. This is also reflected from Fig. 18(d) where FOPID$_4$ structure has the best set-point tracking performance. From Fig. 19(b) and (d), $J_d$ and $J_{track}$ is best for the FOPID$_4$ controller. However the FOPID$_4$ structure is worst in $J_u$ performance. Although it has not been optimized, the peak values of $|S(j\omega)|$ and $|T((j\omega))|$ are lower for the FOPID$_4$ structure as in Fig. 19(a). This indicates that the FOPID$_4$ structure also gives better robust stability and noise rejection performance than the other structures.

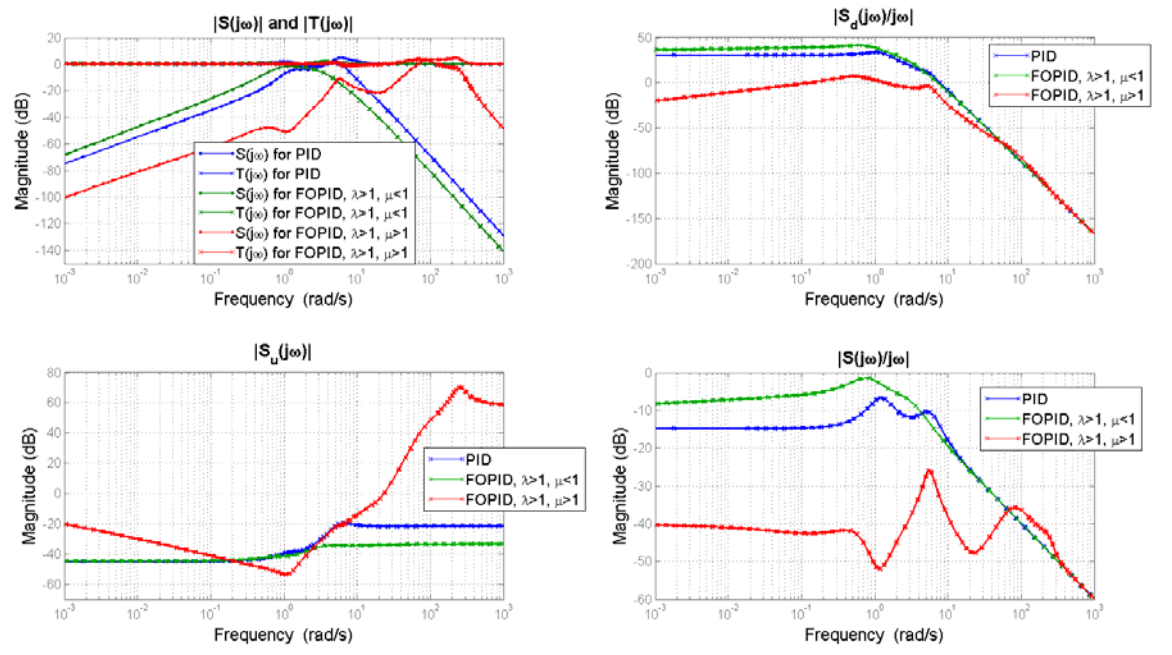

Fig. 20 Best compromise responses for $J_u$-$J_{ST}$-$J_{track}$ trade-off (Case 9).

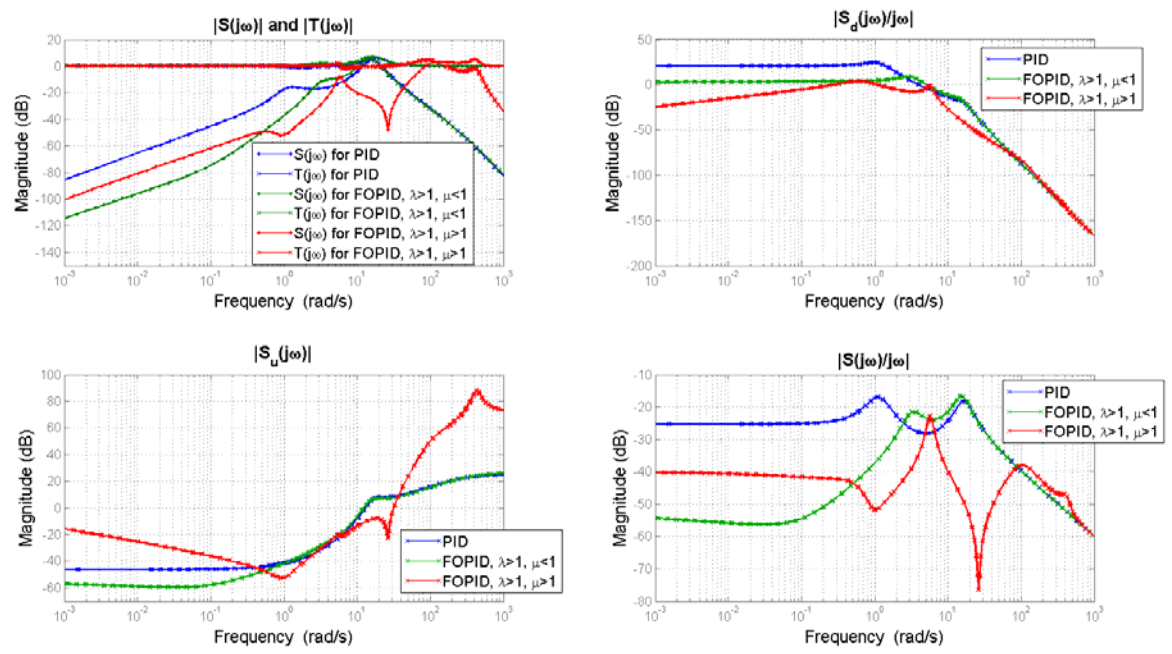

Fig. 21 Best compromise responses for $J_d$-$J_{ST}$-$J_{track}$ trade-off (Case 10).

From Fig. 20(c) and (d), the FOPID$_4$ structure has the best tracking performance but requires the highest amount of control effort. From Fig. 20(a) the FOPID$_4$ structure gives considerably lower values of $\left|S(j\omega)\right|$ and $\left|T(j\omega)\right|$ (indicating more robust stability and noise rejection respectively), than the other two structures. Although not taken explicitly into the objective function, the $FOPID_4$ structure is also the best in disturbance rejection performance as indicated by Fig. 20(b). From Fig. 21(b) and (d), the $FOPID_4$ structure has the best disturbance rejection performance and better overall tracking performance. However, as observed from the plots of $\left|S(j\omega)\right|$ in Fig. 21(a), the tracking performance for the $FOPID_2$ structure is better at lower frequencies. In Fig. 21(a), the peak values of $\left|S(j\omega)\right|$ and $\left|T((j\omega))\right|$ are lower for the $FOPID_4$ structure. The control effort indicated by Fig. 21(c) has not been taken into account during the optimization process. Also, the $FOPID_4$ structure requires a higher amount of control effort than the others.

Overall, the $FOPID_4$ structure can be seen to outperform the PID and the $FOPID_2$ structures in terms of handling multiple conflicting objectives simultaneously. This implies that an AVR system designed with these characteristics would be able to handle unmodelled generator excitation system dynamics without making the system unstable (due to better robust stability). It would also be able to damp out the sudden disturbances in the alternator excitation control loop and maintain the synchronous generator's terminal voltage at a specified level [6]. For any sudden change in the power generation, the FOPID based AVR would be able to reach the specified terminal voltage set-point much faster and without too much undesirable oscillations. It would also be able to eliminate undesirable noise components (due to lower peaks of complementary sensitivity function). However, in most cases, the $FOPID_4$ structure requires a higher amount of control signal to achieve these performance improvements. From the power system perspective this implies that the actuator sizing i.e. the input to the generator excitation system must be higher so as to avoid actuator saturation.

### *D. Performance Comparison with Gain-Phase Margin (GPM) Design*

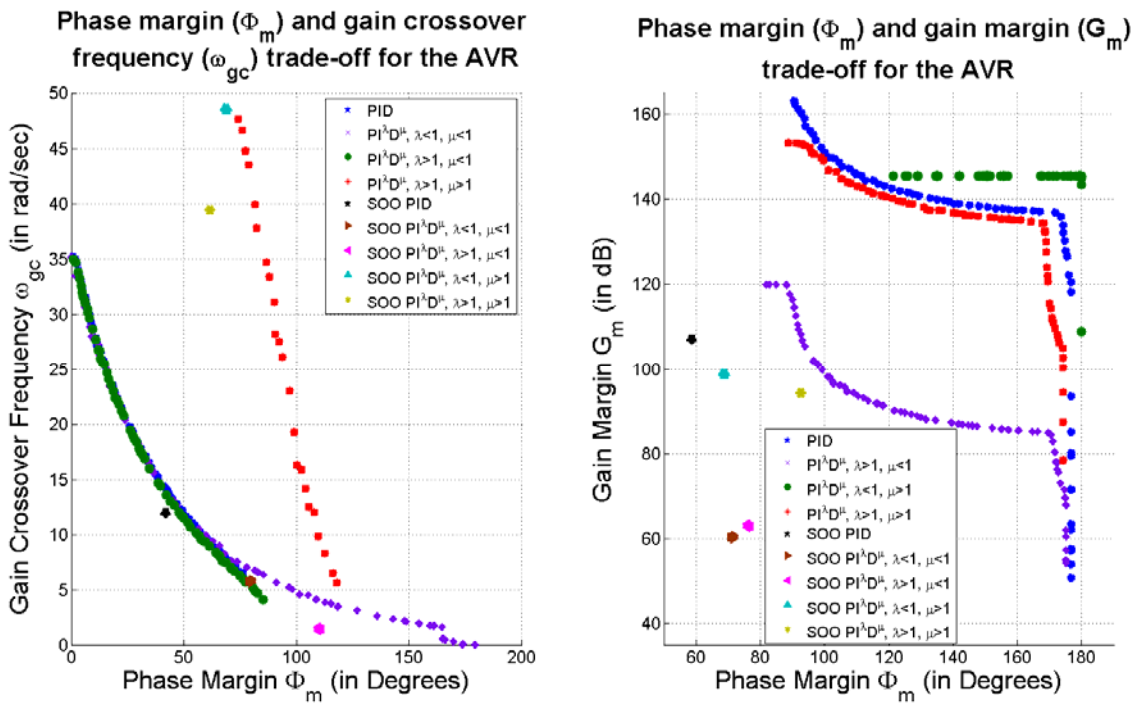

Fig. 22 Frequency domain GPM design trade-off among two objectives: $\Phi_m$-$\omega_{gc}$ (Case 11) and $\Phi_m$-$G_m$ (Case 12).

In order to highlight the advantages of mixed $H_2/H_\infty$ trade-off design, a performance comparison is made with four different types of controller design procedures based on the well-known GPM method of PID controller design [29]. The achievable performance bounds for the GPM method for FO control are explored in [31, 32]. Here, two multi-objective design tradeoffs between $\Phi_m$ - $\omega_{gc}$ (Case 11) and $\Phi_m$ - $G_m$ (Case 12) are considered [9]. The task of the MOO employed here is to maximize for the tradeoffs i.e. $\Phi_m$ vs. $\omega_{gc}$ and $\Phi_m$ vs. $G_m$, based on equation (19) along with constraints imposed for guaranteed stability i.e. gain and phase margin both being positive; gain margin and phase crossover frequency ($\omega_{pc}$) not being infinite and $\omega_{pc} > \omega_{gc}$.

$$\begin{aligned} &Arg\left[L(j\omega_{gc})\right] = Arg\left[C(j\omega_{gc})G_{eff}(j\omega_{gc})\right] = -\pi + \Phi_m \\ &\left|L(j\omega_{gc})\right| = \left|C(j\omega_{gc})G_{eff}(j\omega_{gc})\right| = 1 \\ &\left|L(j\omega_{pc})\right| = \left|C(j\omega_{pc})G_{eff}(j\omega_{pc})\right| = 1/G_m \\ &Arg\left[L(j\omega_{pc})\right] = Arg\left[C(j\omega_{pc})G_{eff}(j\omega_{pc})\right] = -\pi \end{aligned} \quad (19)$$

The other two methods used for comparison are single objective versions of the GPM method which was first done by Monje *et al.* [32] in the context of FO control within a single objective optimization (SOO) framework, to assign $\Phi_m$, $\omega_{gc}$ and ensure iso-damping. The maximization of $\Phi_m$ and $\omega_{gc}$ and $\Phi_m$ and $G_m$ in a SOO framework (Fig. 22) shows that the single objective solutions (with equal weights on both the objectives) are completely dominated by the respective MOO solutions for each controller structure. Also the $FOPID_4$ in case 11 and $FOPID_3$ in case 12 dominate the PID solutions. The $H_2/H_\infty$ performance measures of the fuzzy logic based best compromise solutions of these two design trade-offs are also investigated in Fig. 23-24 respectively.

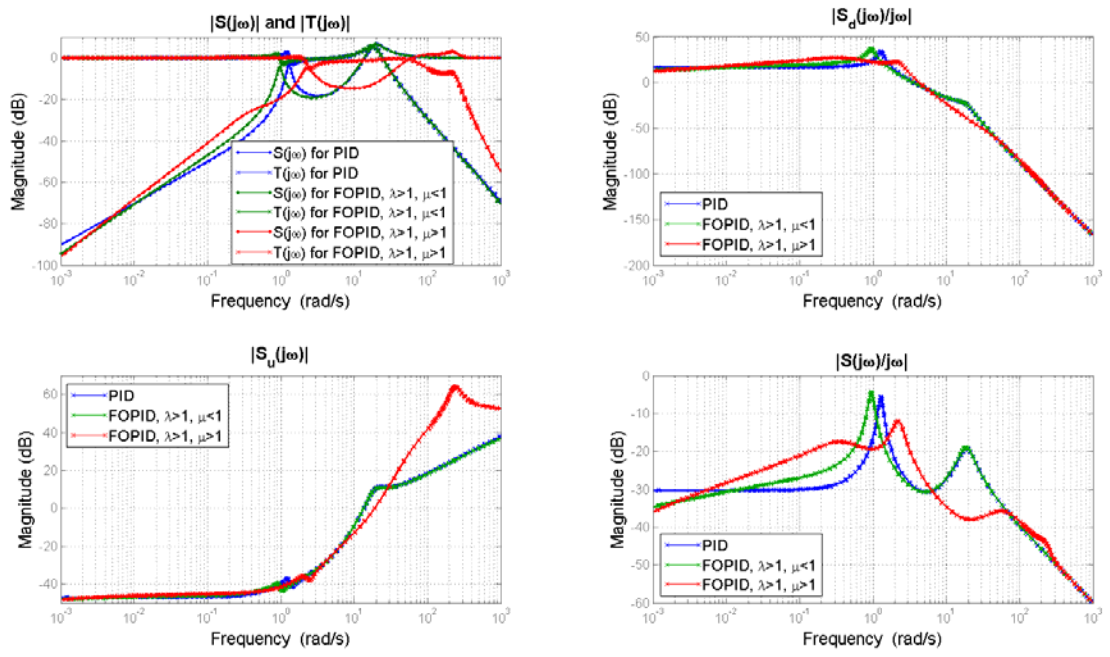

Fig. 23 Best compromise responses for $\Phi_m$-$\omega_{gc}$ trade-off (Case 11).

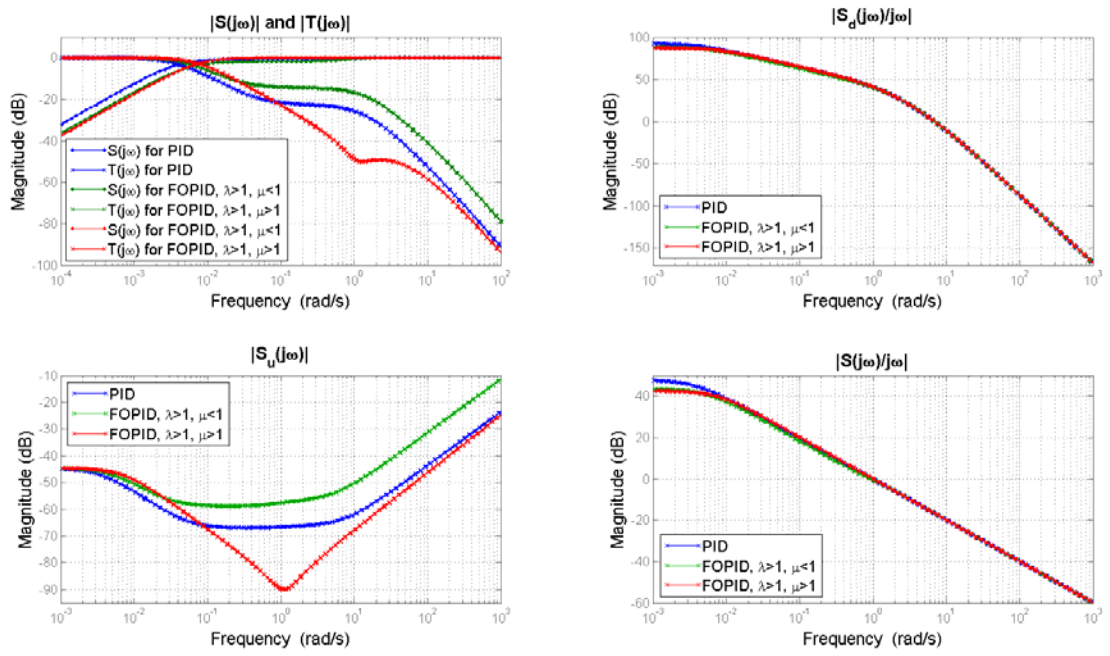

Fig. 24 Best compromise responses for $\Phi_m$-$G_m$ trade-off (Case 12).

It is evident that when the system $H_2/H_\infty$ norms are explicitly taken into consideration, the performance of the best compromise solution is much better. It is also observed that SOO of $\Phi_m$-$\omega_{gc}$ and $\Phi_m$-$G_m$ cannot enforce robust stability, disturbance rejection, small control signal and fast tracking performance which is otherwise possible in the proposed first 10 cases of mixed $H_2/H_\infty$ trade-off designs. It is evident from Fig. 23-24 that the GPM trade-off design suffers from poor disturbance rejection characteristics and high control signal (also evident from Table 1 in the Supplementary material).

## IV. Conclusions

This paper investigated the frequency domain design trade-offs for an AVR system, using a FO-$PI^{\lambda}D^{\mu}$ controller and a multi-objective optimization framework. Due to their greater degree of flexibility, FOPID controllers are able to simultaneously satisfy multiple conflicting design objectives to a greater extent than the simple PID structure which are traditionally used in the industry. The $FOPID_4$ controller structure with $\{\lambda,\mu\}>1$ is found to outperform the others for most design cases. A comparison of the proposed method with four other controller design procedures based on GPM method is also done to highlight the effectiveness of the present approach. Future work can be done on designing FO controllers with nonlinear effects in the AVR control loop.

# Supplementary Material

TABLE I
FUZZY LOGIC BASED BEST COMPROMISE SOLUTIONS FOR DIFFERENT DESIGN TRADE-OFFS AND RESPECTIVE CONTROLLER PARAMETERS

| Trade-off number | Controller Structure | Objective functions | | | | Controller gains | | | | | |
|---|---|---|---|---|---|---|---|---|---|---|---|
| | | $J_d$ | $J_{ST}$ | $J_{track}$ | $J_u$ | $K_p$ | $K_i$ | $K_d$ | $T_f$ | $\lambda$ | $\mu$ |
| 1 | PID | 2.0413 | 0.8449 | - | - | 0.16736 | 0.64860 | 0.03387 | 0.00062 | - | - |
| | $FOPID_2$ | 1.3446 | 0.8996 | - | - | 0.22981 | 1.05204 | 0.04274 | 0.00030 | 1.02388 | 0.99819 |
| | $FOPID_4$ | 2.9257 | 0.7781 | - | - | 0.17148 | 0.34768 | 0.01163 | 0.01920 | 1.08619 | 1.32913 |
| 2 | PID | 1.4947 | - | 0.2862 | - | 0.22325 | 1.41079 | 0.10224 | 0.00006 | - | - |
| | $FOPID_2$ | 1.0765 | - | 0.3014 | - | 0.27722 | 1.78206 | 0.10294 | 0.00008 | 1.00123 | 0.99907 |
| | $FOPID_4$ | 0.3085 | - | 0.1024 | - | 3.66876 | 3.56893 | 0.05366 | 0.00002 | 1.38923 | 1.77811 |
| 3 | PID | 6.0668 | - | - | 0.3491 | 0.07098 | 0.21113 | 0.01414 | 0.05652 | - | - |
| | $FOPID_2$ | 6.4869 | - | - | 0.4051 | 0.05027 | 0.24080 | 0.02527 | 0.07952 | 1.03402 | 0.77364 |
| | $FOPID_4$ | 0.5163 | - | - | 4.5370 | 0.54962 | 2.27603 | 0.03127 | 0.00839 | 1.03814 | 1.23752 |
| 4 | PID | - | 0.6980 | 0.3888 | - | 0.03274 | 0.03316 | 0.01235 | 0.00174 | - | - |
| | $FOPID_2$ | - | 0.6910 | 0.4086 | - | 0.03181 | 0.02779 | 0.01095 | 0.00235 | 1.02931 | 0.97011 |
| | $FOPID_4$ | - | 0.1165 | 0.1203 | - | 1.08832 | 0.26349 | 0.05783 | 0.00006 | 1.57625 | 1.76203 |
| 5 | PID | - | 0.6376 | - | 0.0058 | 0.00344 | 0.00355 | 0.00001 | 0.87218 | - | - |
| | $FOPID_2$ | - | 0.6375 | - | 0.0059 | 0.00134 | 0.00243 | 0.00004 | 1.22557 | 1.00019 | 0.79441 |
| | $FOPID_4$ | - | 0.2208 | - | 564.0656 | 0.07369 | 0.01748 | 0.05572 | 0.00029 | 1.44578 | 1.65018 |
| 6 | PID | - | - | 0.2584 | 123.8108 | 0.06576 | 0.15121 | 0.08178 | 0.00066 | - | - |
| | $FOPID_2$ | - | - | 0.2703 | 23.0614 | 0.06738 | 0.14761 | 0.07520 | 0.00317 | 1.00595 | 0.99997 |
| | $FOPID_4$ | - | - | 0.1446 | 225.7610 | 0.61998 | 0.21681 | 0.04230 | 0.00032 | 1.59033 | 1.56197 |
| 7 | PID | 10.5974 | 0.7242 | - | 0.2951 | 0.04617 | 0.15251 | 0.01527 | 0.06133 | - | - |
| | $FOPID_2$ | 91.1058 | 0.6508 | - | 0.0162 | 0.00245 | 0.01158 | 0.02428 | 2.54075 | 1.03605 | 0.35031 |
| | $FOPID_4$ | 1.7149 | 0.3745 | - | 265.4168 | 0.57796 | 1.08846 | 0.03805 | 0.00038 | 1.34777 | 1.64559 |
| 8 | PID | 1.0495 | - | 0.3222 | 485.0246 | 0.35298 | 1.03791 | 0.08675 | 0.00018 | - | - |
| | $FOPID_2$ | 7.3254 | - | 0.2813 | 22.3467 | 0.08438 | 0.30675 | 0.08970 | 0.00393 | 1.00966 | 0.99988 |
| | $FOPID_4$ | 0.5964 | - | 0.1486 | 501.9223 | 1.55567 | 1.54129 | 0.03338 | 0.00022 | 1.18960 | 1.70488 |
| 9 | PID | - | 0.6815 | 0.5120 | 0.1029 | 0.01373 | 0.03074 | 0.01246 | 0.18237 | - | - |
| | $FOPID_2$ | - | 0.6454 | 0.7118 | 0.0213 | 0.00505 | 0.01037 | 0.00767 | 0.47019 | 1.05315 | 0.70102 |
| | $FOPID_4$ | - | 0.1620 | 0.1111 | 3219.3000 | 0.97572 | 0.41926 | 0.04109 | 0.00005 | 1.46161 | 1.77477 |
| 10 | PID | 16.2302 | 0.9290 | 0.2715 | - | 0.05548 | 0.09149 | 0.06325 | 0.00350 | - | - |
| | $FOPID_2$ | 2.7332 | 0.9218 | 0.2972 | - | 0.14796 | 0.64427 | 0.05958 | 0.00279 | 1.01723 | 0.99589 |
| | $FOPID_4$ | 1.5262 | 0.0862 | 0.1000 | - | 1.27345 | 0.66635 | 0.04448 | 0.00001 | 1.46676 | 1.86088 |
| 11 | PID | 48.7582 | 0.9970 | 0.2841 | 348.0719 | 0.01543 | 0.14547 | 0.08077 | 0.00023 | - | - |
| | $FOPID_2$ | 69.8134 | 0.9863 | 0.2947 | 139.2903 | 0.03051 | 0.08119 | 0.07790 | 0.00046 | 1.13694 | 0.99992 |
| | $FOPID_4$ | 22.8839 | 0.3199 | 0.1973 | 1583.5000 | 0.10711 | 0.03393 | 0.02066 | 0.00005 | 1.28477 | 1.79596 |
| 12 | PID | 41841.0000 | 0.6377 | 10.7061 | 206.3096 | 0.00047 | 0.00002 | 0.00007 | 0.00000 | - | - |
| | $FOPID_2$ | 25630.0000 | 0.6382 | 7.8955 | 2.4230 | 0.00139 | 0.00004 | 0.00027 | 0.00001 | 1.00372 | 0.99998 |
| | $FOPID_4$ | 22446.0000 | 0.6376 | 8.2693 | 2.2829 | 0.00004 | 0.00004 | 0.00003 | 0.00002 | 1.01080 | 1.07051 |